# Mixed-Variable Requirements Roadmaps and their Role in the Requirements Engineering of Adaptive Systems

Ivan J. Jureta
*University of Namur*
ivan.jureta@fundp.ac.be

Alexander Borgida
*Rutgers University*
borgida@cs.rutgers.edu

Neil A. Ernst
*University of Toronto*
nernst@cs.toronto.edu

*Abstract*—The requirements roadmap concept is introduced as a solution to the problem of the requirements engineering of adaptive systems. The concept requires a new general definition of the requirements problem which allows for quantitative (numeric) variables, together with qualitative (binary boolean) propositional variables, and distinguishes monitored from controlled variables for use in control loops. We study the consequences of these changes, and argue that the requirements roadmap concept bridges the gap between current general definitions of the requirements problem and its notion of solution, and the research into the relaxation of requirements, the evaluation of their partial satisfaction, and the monitoring and control of requirements, all topics of particular interest in the engineering of requirements for adaptive systems [3]. From the theoretical perspective, we show clearly and formally the fundamental differences between more traditional conception of requirements engineering (e.g., Zave & Jackson [16]) and the requirements engineering of adaptive systems (from Fickas & Feather [6], over Letier & van Lamsweerde [9], and up to Whittle et al. [14] and the most recent research). From the engineering perspective, we define a proto-framework for early requirements engineering of adaptive systems, which illustrates the features needed in future requirements frameworks for this class of systems.

*Keywords*-requirements roadmap, requirements problem, adaptive systems, relaxation of requirements, monitoring

## I. INTRODUCTION

There is a growing interest in the requirements engineering (RE) of adaptive systems, whose salient feature is the ability to adjust behavior in response to requirements and environment variations. The robustness and versatility that these systems promise will likely result in future sustained investment in theoretical and applied RE research.

The salient feature of research in RE of adaptive systems is disconnectedness (see Cheng et al. [3] for a general overview, Whittle et al. [14] for a recent and comprehensive list of references in RE). This reflects the very novelty of this class of systems, so that mature results were not to be expected. But disconnectedness causes a number of very practical problems which should be discussed already now, reflected by such general questions as "How *exactly* is RE for this class of systems different from established RE?", "How do the inputs and outputs of RE for adaptive systems differ from those in established RE?", as well as more specific ones, e.g., "What should we keep, remove from, or add to existing RE modeling languages/frameworks/methodologies to use them for RE of adaptive systems?".

The aim of this paper is to give clear and formal answers to these questions, and thereby help connect existing research, inform future work, move closer to a deeper classification of research in RE of adaptive systems, and help skeptics decide if/how novel is the problem posed to RE by adaptive systems.

To reach this aim, we illustrate and define the requirements problem for adaptive systems, using a formalism which is just expressive enough to include concepts and relations already appearing in RE, and which includes features central for RE of adaptive systems: (i) monitoring and control of requirements (following Fickas & Feather [6], Feather et al. [5], and Robinson [11]); (ii) probabilistic relaxation of requirements (Letier & van Lamsweerde [9]); (iii) fuzzy relaxation of requirements (Whittle et al. [14] and Baresi et al. [2]); and (iv) adaptation requirements (Souza et al. [13]). It turns out that extending the simple, but abstract modeling language Techne [7] to allow constraints over quantitative (numerical) variables is enough to capture key ideas in monitoring, relaxation, and adaptation, and to show how they participate in the requirements problem of adaptive systems.

This paper presents two main contributions:

- The requirements problem for adaptive systems and its solution concept, called *requirements roadmap* (roadmap hereafter), are formally defined. A roadmap amounts to a sequence of requirements configurations, each of which is shown to satisfy all properties required by Zave & Jackson [16], and in our prior work [8]. These parallels show exactly how the problem and roadmap concepts are different from more traditional RE.
- To introduce the problem and roadmap concepts, we define a simple formalism in which we can model (i) probabilistic and fuzzy relaxation, (ii) monitored and controlled variables, (iii) adaptation requirements. It is a proto-framework for RE because (a) it is impractical in its current form, but is enough to allow us to present and discuss the salient properties of the problem and roadmap concepts, and (b) it can serve as a prototype for early formal modeling languages for RE of adaptive systems.

We use the London Ambulance Service (LAS) case study [1] to recall and illustrate the key notions from prior work that we mentioned above (§II), then to informally illustrate the new problem and roadmap concepts (§III). We discuss their departure from prior work (§IV), and present the formalization of the concepts and relations used to define these concepts and model there instances throughout the paper (§V). Finally, we summarize conclusions and open issues (§VI).

## II. BASELINE & RELATED WORK

We discuss in this section how earlier contributions in RE – concerning monitoring and control, probabilistic relaxation, and fuzzy relaxation – influence the understanding of the requirements problem and its solutions for adaptive systems.

### A. Background

We assume that requirements are either propositional variables or constraints on quantitative variables, which are categorized according to the core ontology [8], and that relations are formalized as in Techne [7]; a requirement is

- **k**: a domain assumption if it states a value that a variable is believed to have; e.g., in Figure 1(a), $\mathbf{k}(r_1)$ is a domain assumption over a propositional, qualitative variable, while in Figure 1(d), $\mathbf{k}(\mu(t_x) = 0.6)$ is a domain assumption over a quantitative variable, the value of a function $\mu$;
- **g**: a goal if it identifies a desirable value of a qualitative variable, i.e., it states a property that we want to see holding; e.g., $\mathbf{g}(p_2)$ in Figure 1(a);
- **q**: a quality constraint if it places a constraint on desirable values of a quantitative variable; e.g., in Figure 1(d), $\mathbf{q}(t_x = 60\text{sec})$ says that the desirable value of $t_x$ is 60sec;
- **s**: a softgoal if it refers to a desirable value of a variable in such a way that it is not possible to identify exactly which value, or range of values of that variable it refers to; e.g., in Figure 1(d), $\mathbf{s}(w_1)$ is a softgoal.
- **t**: a task if the propositional variable in it says what to do; e.g., in Figure 1(a), $\mathbf{t}(u_1)$ is a task.

A requirement can be mandatory, optional, or neither. $\mathbf{g}(q_2)^\mathbf{M}$ is a mandatory goal, which means that it must be satisfied. $\mathbf{g}(p_{15})^\mathbf{O}$ is an optional goal, meaning that it is more desirable that it is satisfied, than not, but we will still accept a system which fails to satisfy $\mathbf{g}(p_{15})^\mathbf{O}$. Requirements can be in three binary or n-ary relations. The directed arrow in Figure 1(a) represents the *inference* relation [7], also understood as a conditional, if-then relation, so that the line from $\mathbf{t}(u_1)$ to $\mathbf{g}(p_2)$ abbreviates the formula $\mathbf{k}(\mathbf{t}(u_1) \to \mathbf{g}(p_2))$, i.e., a domain assumption according to which, if we can deduce $\mathbf{t}(u_1)$, then we can also deduce $\mathbf{g}(p_2)$, or in other words, $\mathbf{t}(u_1)$ operationalizes $\mathbf{g}(p_2)$. The inference relation stands for operationalization when it goes from tasks and domain assumptions to goals; otherwise, it can be read as a refinement: in Figure 1(a), $\mathbf{k}(\mathbf{k}(r_1) \wedge \mathbf{k}(r_2) \wedge \mathbf{g}(p_1) \wedge \mathbf{g}(p_2) \wedge \mathbf{g}(p_3) \wedge \mathbf{g}(p_4) \to \mathbf{g}(q_2)^\mathbf{M})$ says that the domain assumptions $\mathbf{k}(r_1)$ and $\mathbf{k}(r_2)$, and the goals $\mathbf{g}(p_1)$ to $\mathbf{g}(p_4)$ refine $\mathbf{g}(q_2)^\mathbf{M}$. Logically inconsistent requirements are in the *conflict* relation: e.g., we assume in Figure 1(a) that $\mathbf{t}(u_2)$ cannot be satisfied together with $\mathbf{t}(u_4)$, which we draw with the bulleted line between them, and this abbreviates the assumption $\mathbf{k}(\mathbf{t}(u_2) \wedge \mathbf{t}(u_4) \to \bot)$: if we can deduce both, then we will also conclude logical inconsistency. Finally, we can have a preference relation between requirements: e.g., $\mathbf{t}(u_{10}) \succ \mathbf{t}(u_{13})$ in Figure 1(d) reads that satisfying $\mathbf{t}(u_{10})$ is strictly more desirable than satisfying $\mathbf{t}(u_{13})$.

### B. Monitoring, Control, Probabilistic & Fuzzy Relaxation

Enabling adaptive behavior requires the monitoring of requirements and control of behaviors intended to satisfy requirements. Fickas & Feather [6] suggested that for this we must start with a specification of all alternative behaviors to be considered in a requirments model, that is, alternative refinements and operationalizations of all requirements. The model is then a source of *monitored* variables, the observed values of which will trigger behaviors, called reconciliation tactics, that are intended to change the values of *control* variables, which can affect how one operationalizes a requirement. This forms a *control loop*, through which the system observes and reacts by switching from one runtime configuration to another. For illustration, consider the goal $\mathbf{g}(p_3$ : Check if double location) for LAS in Figure 1(a), which is operationalized via two refinements: one way to satisfy $\mathbf{g}(p_3)$ is by executing both tasks $\mathbf{t}(u_2)$ and $\mathbf{t}(u_3)$, the other is via $\mathbf{t}(u_4)$. A reconciliation tactic could be, e.g., if $\mathbf{t}(u_2)$ fails, then operationalize $\mathbf{g}(p_3)$ by $\mathbf{t}(u_4)$. Feather et al. [5] combined the requirements monitoring mechanisms from Fickas & Feather with KAOS [4] to make an RE framework, where requirements can be converted into constraints on events, and events are monitored at runtime. Robinson [11] suggested a requirements monitoring platform that can be combined with a requirements modeling language via translation rules between the expressions of the latter and those accepted by the former, then suggested how to use OCL as the modeling language for the platform [12].

An important characteristic of a reconciliation tactic (and same applies to adaptation goals [2]), is that it is defined in relation to *local* requirements, i.e., it says what to do in relation to a single requirement which is not being satisfied. There are two problems with defining **adaptation requirements** by looking mainly at local requirements. First, it is not clear that following any or every reconciliation tactic will make sure that the new system configuration is consistent with the requirements. It is not clear what happens if the reconciliation tactic works locally, but activates, e.g., tasks which are inconsistent with some other already active part of the system. The other problem is that one local change may seem fine by itself, and may result in a new and consistent configuration, but this new configuration may be a dominated one: i.e., maybe instead of satisfying one adaptation requirement and obtaining some quality level, we could have used several adaptations, which result in a configuration that ensures higher quality. The local approach to the definition of adaptation requirements comes with the risk of adapting to a sub-optimal or inconsistent configuration.

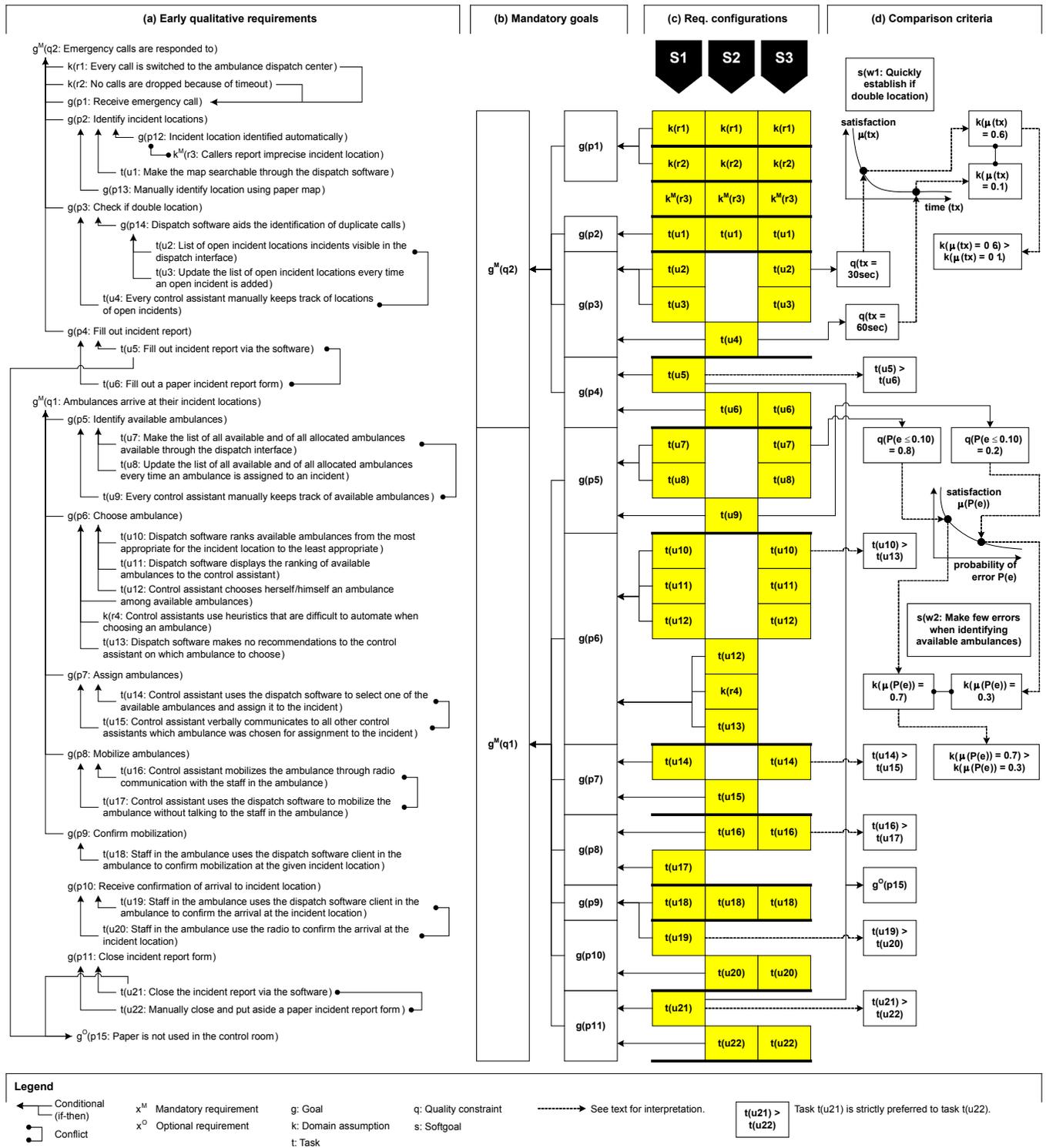

Figure 1. Simple early requirements, requirements configurations, and comparison criteria for LAS.

We therefore contend that instead of defining adaptation requirements by considering their impact only on a small set of requirements, *adaptation requirements should be defined so*

*as to ensure that the configuration obtained after adaptation is desirable in its entirety in terms of various criteria used to compare alternative requirements configurations* (including, for example, preferences over requirements, probabilities of requirements satisfaction, and so on). This has two consequences on the formulation of the problem and solution concepts in adaptive system requirements. First, it follows that we do not define adaptation requirements while writing the requirements model, but *compute* them after we have identified at least some configurations from a set of alternatives, all of which satisfy some set of properties (e.g., are consistent) and satisfy some preference requirements. Consider Figure 1(c), where each column $S_1$, $S_2$, and $S_3$ is a different requirements configuration, and each satisfies the mandatory goals set in Figure 1(a) for the LAS system: e.g., if the system is configured according to $S_1$, and $\mathbf{t}(u_5)$ fails, then it may be more desirable to adapt to $S_3$ than to $S_2$, so that adaptation requirements can be identified directly by comparing $S_1$ to $S_3$, and defined as the switch from $\mathbf{t}(u_{21})$ to $\mathbf{t}(u_{22})$, from $\mathbf{t}(u_{19})$ to $\mathbf{t}(u_{20})$, from $\mathbf{t}(u_{17})$ to $\mathbf{t}(u_{16})$, and from $\mathbf{t}(u_5)$ to $\mathbf{t}(u_6)$. Secondly, the local/global distinction has an impact on how we *select* configurations: it is likely that the system *can* adapt from one to *one among several alternative configurations*; e.g., if $\mathbf{t}(u_{17})$ fails in Figure 1(c), both $S_2$ and $S_3$ use the alternative $\mathbf{t}(u_{18})$. Selection should take into account the consequences of adapting to a configuration: e.g., perhaps configuration $S_1$ is more desirable than $S_2$ if only these two are compared, but we may know that if we select $S_1$, then in the long term, e.g., it is not sustainable, in order to keep satisfying some quality constraint. This means that there can be preferences over sequences of configurations.

Adaptive behavior by definition means that the system cannot satisfy all the specific requirements and meet the desired quality levels, all of the time – it may do worse, but it may also do better. **Relaxation** of requirements has been suggested to cover both cases: a requirement can be relaxed to allow the system to fail it, but we restrict how often it can do so (via probabilistic relaxation), or to avoid overly constraining the system when it may deliver higher quality than expected (via fuzzy relaxation). Relaxation of requirements is closely related to the evaluation of their satifaction, as one aim of relaxation is to quantify the degree of satisfaction of a requirement. Two broad approaches to relaxation and satisfaction evaluation have been suggested, based on *objective criteria* (where quantification is in terms of measurable phenomena in the environment or the system) or on *subjective criteria* (where a number is given to reflect personal impression of satisfaction, regardless of measurable phenomena in the domain).

An important approach to relaxation, based on objective criteria, appears in Letier & van Lamsweerde [9], who extended goal models from KAOS to allow the specification of measures on system behaviors, and probability functions over the values of these measures. Using such a formalism, they can state, e.g., that in the LAS system, 95% of ambulances should reach the designated place of incident in at most 14 minutes after an incident is reported. Values of measures and probabilities are computed for every requirements configuration, all of which are inputs to a decision-making process that aims at selecting a single configuration.

Approaches based on subjective criteria were revived recently for RE of adaptive systems. Whittle et al. [14] associate fuzzy membership functions to qualitative requirements, in order to allow and quantify departures from the ideal satisfaction of such requirements. The fuzzy membership function can be interpreted as a satisfaction function, and they use modalities to relax requirements: e.g., to relax the task $\mathbf{t}(u_{18})$ in Figure 1, we would write [As early as possible]$\mathbf{t}(u_{18})$. This means that there is a function $\mu$ which returns a level of satisfaction when given a duration between time "now" and the time when $\mathbf{t}(u_{18})$ is executed, whereby $\mu$ "has its maximum value at 0 (i.e., at the current point in time) [and] tails off gradually ad infinitum (i.e., it has a triangular membership graph that is asymptotic)" [14].

Baresi et al. [2] also use fuzzy membership functions to treat qualitative requirements as quantitative ones. They introduce degrees of satisfaction between the binary 0 (for violated) and 1 (satisfied); thus, adjusted requirements are associated to "adaptive" requirements that relate ranges of values of the fuzzy membership function to trigger switching between configurations. For example, if there is a quality constraint $\mathbf{q}(v < 6\text{hrs})$ here (a goal $\mathcal{G}(v < 6\text{hrs})$ for them, as they allow quantitative variables in goals), then relaxing it would amount to replacing it with $\mathbf{q}(v <_f 6\text{hrs})$ (in their notation, $\mathcal{G}(v <_f 6\text{hrs})$), where $\leq_f$ is a fuzzy operator. Their interpretation of $v <_f 6\text{hrs}$ is that there is a fuzzy membership function $\mu$ which returns the level of satisfaction as a function of $v$, and the shape of $\mu$ is predefined (for $<_f$ in $\mathcal{G}(v <_f 6\text{hrs})$, it is positive and constant until $v = 6$, then decreases up to the satisfaction value 0 for some $v > 6$).

## III. RPAS & ROADMAP CONCEPTS

The *requirements problem for adaptive systems* (RPAS) and the corresponding roadmap concept need to be introduced because monitoring, control, adaptive requirements, and probabilistic and fuzzy relaxation cannot be formulated in the traditional requirements problem and solution framework: these features add information to the requirements problem, which in turn affects the properties that solutions need to satisfy, and influence the criteria used to compare alternative solutions. The aim of this section is to present and discuss the properties of the roadmap concept, as the solution concept for RPAS; we later compare RPAS and roadmap concepts to Zave & Jackson [16] and Techne [8] (cf., §IV).

### A. Requirements Problem for Adaptive Systems

The requirements problem is a *design* and *decision-making* problem. By design, we mean that alternative solutions to the problem are not available, but need to be created: e.g., in Figure 1(a), we start from two broad goals, $\mathsf{g}(q_1)$ and $\mathsf{g}(q_2)$, so that we need to elicit further requirements and information, refine goals, identify conflicts, define operationalizations, all of which fall under design.[1] Decision-making requires the identification of criteria for the comparison of alternative solutions, and the application of a decision rule to rank alternative solutions from the most desirable to the least desirable.

To provide definitions, we start with the notion of a requirements database $\Delta$ — a set of requirements (i.e., domain assumptions, goals, quality constraints, softgoals, tasks) and relations between them (e.g., conjunction, inference, refinement, conflict, preference). For example, the requirements database for LAS in this paper includes all labeled propositions in Figure 1.

For the remaining definitions, we start at the end (RPAS), and explain one by one its component definitions. Hence definitions unfold recursively.

RPAS is the problem of designing and ranking roadmaps.

**Definition III.1.** *The requirements problem for adaptive systems: Given a requirements database $\Delta$, design, find, and rank roadmaps that can be defined from $\Delta$.*

A roadmap is a sequence of requirements configurations and adaptation requirements needed to switch between these configurations.

**Definition III.2.** *A requirements roadmap is a pair $\mathcal{R} = (\mathcal{S}, \mathcal{A})$, where $\mathcal{S} = (S_1, \ldots, S_n)$ is a sequence of $n$ requirements configurations and $\mathcal{A}$ is a set of adaptation requirements.*

The novelty of the problem and roadmap concepts lies in the properties of requirements configurations, adaptation requirements, and in how roadmaps can be ranked. We discuss these in turn below.

*B. Properties of Requirements Configurations*

Let us first look at the configurations appearing in a roadmap.

**Definition III.3.** *A configuration in a roadmap is a set $S$ of domain assumptions and tasks in $\Delta$ such that $S$ satisfies the Consistency, Qualitative threshold achievement, Quantitative threshold achievement, Conformity, Dominance, and Minimality conditions defined below.*

*1) Consistency:* Every requirements configuration $S$ must be consistent, i.e., $S \not\vdash_{\widehat{\tau}} \bot$, where $\vdash_{\widehat{\tau}}$ is the Techne consequence relation [7]. With consistency, the adaptive system should work to satisfy a consistent set of requirements. In Figure 1(c), every configuration is such that there are no conflicts between its members. Remark that given a sequence $\mathcal{S} = (S_1, \ldots, S_n)$ in a road map, $\bigcup \mathcal{S}$ can, and will often be inconsistent, as consecutive configurations need not be consistent together. If we adapt from $S_1$ to $S_2$ in Figure 1(c), note that $S_1 \cup S_2 \vdash_{\widehat{\tau}} \bot$, since there are conflicts, e.g., between $\mathsf{t}(u_{16})$ and $\mathsf{t}(u_{17})$.

*2) Qualitative Threshold Achievement:* Every configuration $S$ should operationalize every mandatory goal in $\Delta$ — a set written as $\Delta_{\mathsf{g}}^{\mathsf{M}}$. An operationlization defines tasks and domain assumptions from which we can deduce (via $\vdash_{\widehat{\tau}}$) the goal. For example, in Figure 1(a), if the task $\mathsf{t}(u_1)$ is satisfied, then $\mathsf{g}(p_2)$ is satisfied, so that the former operationalizes the latter. Since a requirement can be operationalized by alternative sets of tasks and domain assumptions, we have a function $\mathsf{Op}(\varphi)$ which, which given a mandatory goal $\varphi \in \Delta_{\mathsf{g}}^{\mathsf{M}}$, returns the set of all operationalizations of $\varphi$. E.g., $\mathsf{Op}(\mathsf{g}(p_7))$ includes two operationalizations of $\mathsf{g}(p_7)$ in Figure 1(a). The property is restricted to *qualitative* requirements, for which satisfaction is binary, and thus to goals, since a goal is either satisfied or not. The property requires *threshold* satisfaction, since it requires only that all *mandatory* goals be satisfied by a configuration.

A consistent set of requirements must meet both the qualitative and quantitative thresholds, for it otherwise fails to satisfy the requirements which must be satisfied.

*3) Quantitative Threshold Achievement:* Every configuration $S$ should operationalize every mandatory quality constraint in $\Delta$ — a set written as $\Delta_{\mathsf{q}}^{\mathsf{M}}$. Once again, the function $\mathsf{Ov}(\varphi)$, returns all sets of operationalizations of quality constraint $\varphi = \mathsf{q}(\alpha)$, whereby some set $\Pi$ is such an operationalization iff it assigns values to variables in $\alpha$ which ensure that the constraint in $\alpha$ is satisfied. Quality constraints identify desirable values of quantitative variables. E.g., $\mathsf{q}$s may thus place constraints on values of measures of some behavior of the system-to-be. In LAS, and in relation to the response to emergency calls and the mobilization of ambulances, the time that these activities take is a crucial part of quality of service. We can use the following variables, where $c$ identifies a call and $e$ a unique incident:

$t_{1,c}$: Time the caller waits for a control assistant;
$t_{2,c}$: Time to identify incident location;
$t_{3,c}$: Time to fill out incident report;
$t_{4,e}$: Time to mobilize an ambulance;
$t_{5,e}$: Time for the mobilized ambulance to arrive and confirm arrival at incident location.

Instead of defining bounds on all of these variables, a government standard may not be as specific, and may impose a maximal time over a subset of activities that need to be executed between the placement of the emergency call and the confirmation of arrival of an ambulance to location; e.g.:

$\mathsf{q}(t_{6,c} \leq 3\text{min}$: $t_{6,c}$ is the duration between the switching of the call $c$ to the dispatch center to the mobilization of the ambulance to the incident location;)

---

[1] Note that "design" is typically used for steps of systems development that come after RE, such as architectural design. The term "design" is used here in a broader sense, to encompass all activities (e.g., elicitation, validation) which are necessary to make alternative options on which we then need to decide.

where it is clear that the value of $t_{6,c}$ depends on $t_{1,c}$, $t_{2,c}$, $t_{3,c}$ and $t_{4,e}$. If we assume that $t_{6,c} = t_{1,c} + t_{2,c} + t_{3,c} + t_{4,e}$, then we add this to the requirements database as a domain assumption over quantitative variables, $\mathbf{k}(t_{6,c} = t_{1,c} + t_{2,c} + t_{3,c} + t_{4,e})$. The domain assumption says that there is a *quantitative refinement* relation between variables, and it specifies the functional relation between the refined $t_{6,c}$ and the variables refining it. In contrast to the refinement relation, quantitative refinement is not over requirements, but variables in requirements.

While it may be useful to set a precise bound on the value of an aggregate variable, as $t_{6,c}$ above, it may be more interesting to set bounds relative to the values of other variables. E.g., if we want $t_{2,c}$ to be at most 110% of its average value over the past three months, then we write

$\mathbf{q}(t_{2,c} \leq (1.1/3)(v_{1,m-3} + v_{1,m-2} + v_{1,m-1}))$;
$\mathbf{k}(v_{1,m} = n(m)^{-1} \sum_{i=1}^{n(m)} \sum_{c=1}^{c(i)} t_{2,c})$, where $m$ is the month identifier, $n(m)$ the number of days in $m$, $c$ is the call identifier on a given day, $c(i)$ is the total number of calls received on day $i$ of month $m$;

where the domain assumption is a quantitative refinement of $v_{1,m}$, the average time, over a month, that it takes to identify the location.

The quality constraints and quantitative refinements need to be related to goals, tasks and domain assumptions in order to determine if the former are satisfied. If there is a running system, the values of $t_{2,c}$ are recorded, and it is straightforward to check if the constraint $t_{2,c} \leq (1.1/3)(v_{1,m-3} + v_{1,m-2} + v_{1,m-1})$ is satisfied. Before the system is in operation, we can simulate values of $t_{2,c}$, by assuming that $t_{2,c}$ is a random variable which has some probability distribution, so that we would have $\mathbf{k}(t_{2,c} \backsim \mathcal{N}(60sec, 45sec^2))$ if we assume that $t_{2,c}$ follows a normal distribution with mean 60sec and variance 45sec² when the task $\mathbf{t}(u_1)$ is satisfied, which we model by a refinement $\mathbf{k}(\mathbf{t}(u_1) \rightarrow \mathbf{k}(t_{2,c} \backsim \mathcal{N}(60sec, 45sec^2)))$. This assumption may be based on data from a pilot study, from expert opinion, or from data on systems which also satisfy $\mathbf{t}(u_1)$ and are already in operation.

*4) Conformity: Every configuration $S$ must include all strict domain assumptions and all mandatory tasks*, i.e., $\Delta_{\mathbf{k}}^{\mathbf{M}} \cup \Delta_{\mathbf{t}}^{\mathbf{M}} \subseteq S$. This property asks that all strict domain assumptions are not violated and all mandatory tasks are executed in every configuration.

The Qualitative and Quantitative threshold achievement and Conformity properties ensure that a configuration satisfies all that *must* be satisfied.

*5) Dominance: Every configuration $S$ must be maximal w.r.t. optional requirements*, i.e., $\nexists S' \subseteq \Delta$ such that (i) both satisfy the Consistency, Qualitative threshold achievement, Quantitative threshold achievement, and Conformity conditions, (ii) $S \subset S'$, and (iii) $S' \setminus S$ contains optional $\mathbf{k}$ or $\mathbf{t}$ requirements.

This condition formalizes the idea of optional requirements, which are desirable to satisfy, but can be violated. A configuration should include as many such defeasible domain assumptions and as many optional tasks, up to the point at which adding any further defeasible domain assumptions and/or optional tasks violates the Consistency, Qualitative and Quantitative threshold achievement, and Conformity properties. The Dominance property ensures that every configuration is Pareto efficient with regards to optional requirements, as this condition makes it impossible to add optional domain assumptions and tasks to any $S$ and still ensure that $S$ is a configuration.

*6) Minimality: A set $S$ satisfying all the above properties must be minimal in order to qualify as a configuration.* Minimality requires that a configuration includes only the domain assumptions and tasks which are needed to satisfy exactly the Consistency, Qualitative threshold achievement, Quantitative threshold achievement, Conformity, and Dominance properties.

Configurations in Figure 1(c) satisfy all six properties.

*C. Adaptation Requirements, Monitoring & Control*

An adaptation requirement is meant to constrain successive configurations in a roadmap. It does so by describing the presence/absence of certain requirements in pairs of related configurations, and is triggered by the failure of monitored assumptions, tasks, or quality constraints. For example, in Figure 1(c), if task $\mathbf{t}(u_{21})$ fails, we might want to replace $\{\mathbf{t}(u_{19}), \mathbf{t}(u_{17}), \mathbf{t}(u_5)\}$ with $\{\mathbf{t}(u_{22}), \mathbf{t}(u_{20}), \mathbf{t}(u_{16}), \mathbf{t}(u_6)\}$. In general, we will therefore want to specify the conditions monitored for failure, and the changes (additions, removals) that must be seen in the successor configuration when failure occurs. The above adaptation requirement can then tell us that if we are running the configuration $S_1$ in Figure 1(c), and the system fails to satisfy $\mathbf{t}(u_{21})$, then it could adapt from $S_1$ to $S_3$. Note that an adaptation requirement $\langle T, A, D \rangle$ might then be thought of as a a simple operator in AI planning, where $T$ is trigger condition, while $A$ and $D$ are add/delete lists.

**Definition III.4.** *An **adaptation requirement** is an operator of the form $\langle T, A, D \rangle$, where $T$ is a set of requirements present in the initial configuration (which are monitored to fail) while $A$ and $D$ are, respectively, sets of requirements that must be present and absent in the final configuration, if the operator applies.*

Note therefore that if operator $\langle T, A, D \rangle$ applied in going from configuration $S_i$ to $S_{i+1}$, then $T \cup D \subseteq S_i$ and $A \cap S_i = $ must hold, and analogously with $S_{i+1}$. Note also that it is quite acceptable that multiple operators apply at the same time, when connecting $S_i$ to $S_{i+1}$, and that (as with the so-called ramification problem in AI), there may be additional changes needed in $S_{i+1}$ in order to make it a valid configuration.

In the above sense, requirements are monitored, while adaptation requirements act as a control mechanism, which guides adaptation by changing the target requirements configuration that the adaptive system needs to satisfy.

*D. Comparison Criteria & Ranking of Roadmaps*

Ranking of roadmaps requires the identification of comparison criteria, and the application of a decision rule which establishes a ranking on the basis of criteria. Comparison criteria are either optional requirements or preferences. Preferences can be individually added to $\Delta$, or can be obtained through the relaxation of requirements, or from softgoals. Preferences are illustrated in Figure 1(d), where, e.g., $\mathbf{t}(u_{16})$ is strictly preferred to $\mathbf{t}(u_{17})$, which indicates that, for this criterion only, a configuration having the former task is strictly more desirable than the configuration having the latter task.

*1) Preferences through Probabilistic Relaxation:* Continuing the earlier example (cf., §III-B3), the upper bound on $t_{2,c}$ in $\mathbf{q}(t_{2,c} \leq (1.1/3)(v_{1,m-3} + v_{1,m-2} + v_{1,m-1}))$ may still be too idealistic, as callers may provide information of very different quality about the incident location.

Probabilistic relaxation of a quality constraint is done in two steps. Firstly, the variable constrained in $\mathbf{q}$ is redefined as a random variable, and an assumption is made on the probability distribution of that random variable. Adding $\mathbf{k}(t_{2,c} \backsim \mathcal{N}(60sec, 45sec^2))$ to $\Delta$ makes of $t_{2,c}$ a random variable which follows a normal distribution. Secondly, the quality constraint to be relaxed is removed from $\Delta$, and a new quality constraint is added. The new constraint specifies a bound not on the value of the now random variable, but on the probability that its value is in some range. Since we made $t_{2,c}$ into a random variable in the first step, we now replace $\mathbf{q}(t_{2,c} \leq (1.1/3)(v_{1,m-3} + v_{1,m-2} + v_{1,m-1}))$ with $\mathbf{q}(P(t_{2,c} \leq (1.1/3)(v_{1,m-3} + v_{1,m-2} + v_{1,m-1})) \geq 0.90)$, that is, we now require that the minimal probability should be 0.90 for $t_{2,c}$ to be at most 110% of its three-month average.

Random variables and quality constraints thereon play an important role in decision-making, as we use them to set desired probability levels over random variables. In more informal terms, this means that we can write quality constraints and domain assumptions which reflect, respectively, the confidence that we desire to reach in relation to system behaviors, and confidence that we actually have. Figures 1(c)–(d) indicate that $\mathbf{t}(u_7)$ operationalizes the quality constraint $\mathbf{q}(P(e \leq 0.10) = 0.8)$, while $\mathbf{t}(u_8)$ operationalizes $\mathbf{q}(P(e \leq 0.10) = 0.2)$. The two quality constraints suggest that the two tasks result in different probabilities of having less than 10% of erroneous identifications of available ambulances. Figure 1(d) further indicates the level of satisfaction with each of the two probability levels.

*2) Preferences through Fuzzy Relaxation:* Fuzzy relaxation of a quality constraint involves two steps. In the first step, the quality constraint is removed, and a fuzzy membership function is defined over the variable from the removed quality constraint. In the second step, a softgoal is defined over the variable from the relaxed requirement. Consider again $\mathbf{q}(t_{2,c} \leq (1.1/3)(v_{1,m-3} + v_{1,m-2} + v_{1,m-1}))$ and we now apply fuzzy relaxation. We start by removing this quality constraint from $\Delta$. A fuzzy membership function over $t_{2,c}$, denote it $\mu(t_{2,c})$, depends on how stakeholders evaluate, in terms of desirability, the various values of $t_{2,c}$. E.g., we could use $\mu(t_{2,c}) = e^{-t_{2,c}}$, so that the higher the value of $t_{2,c}$, the lower $\mu(t_{2,c})$ is, which reflect the idea that the more time it takes to identify an incident location, the more the stakeholders are dissatisfied, whereby their satisfaction increases as $t_{2,c}$ approaches 0 (another example of a satisfaction function is in the upper part of Figure 1(d)). If we adopt $\mu$ as defined, we have removed the quality constraint on $t_{2,c}$ and we quantify satisfaction as a function of $t_{2,c}$.

To finish with the fuzzy relaxation of the quality constraint on $t_{2,c}$, a softgoal needs to be added to the requirements database over values of $t_{2,c}$. Softgoals have had various definitions in RE, but there seems to be agreement on their two properties: (i) they are used for the comparison of alternatives, and (ii) they are vague, as they refer to some desirable values of variables, even though it may not be clear which exact values, or of which variables. E.g., *Response time should be low* is a typical softgoal, where the variable is suggested, but it is not clear which specific range of its values qualify as *low*; in the softgoal *High safety*, not only is it not clear how to measure *safety*, it is also not apparent when *safety* is *high*. When used in fuzzy relaxation, a softgoal is defined over a known variable: in our example, it is reasonable to prefer lower over higher values of $t_{2,c}$, and we consequently have the softgoal

$\mathbf{s}$( LET $x_1 \stackrel{\text{def}}{=} \text{VAL}(S_1, t_{2,c})$ AND $x_2 \stackrel{\text{def}}{=} \text{VAL}(S_2, t_{2,c})$;
   IF $\mu(x_1) > \mu(x_2)$ THEN ADD
   $\{\mathbf{k}(t_{2,c} = x_1), \mathbf{k}(t_{2,c} = x_2),$
   $\mathbf{k}(t_{2,c} = x_1) \succ \mathbf{k}(t_{2,c} = x_2)\}$
   TO $\Delta$; )

where $\text{VAL}(S_1, t_{2,c})$ returns the value of $t_{2,c}$ in $S_1$. Although the formulation above seems very different from saying "Low $t_{2,c}$", this is precisely what it does, as long as $\mu(t_{2,c})$ decreases as $t_{2,c}$ increases. $\succ$ denotes the strict preference relation, and $\mathbf{k}(t_{2,c} = x_1) \succ \mathbf{k}(t_{2,c} = x_2)$ says that satisfying $\mathbf{k}(t_{2,c} = x_1)$ is strictly more desirable than satisfying $\mathbf{k}(t_{2,c} = x_2)$. The important point to see is that the softgoal specifies a macro which generates domain assumptions and preference relations. For any two configurations $S_1$ and $S_2$, the macro compares the satisfaction (returned by satisfaction function $\mu$) with values that $t_{2,c}$ has in each of those two configurations. Depending on the comparison between these values, the macro adds two domain assumptions and a preference relation to $\Delta$. The reason we add them to $\Delta$ and not individual configurations is because we can add one of these domain assumptions only if doing so does not violate the conditions that a configuration must satisfy (i.e., adding one of these domain assumptions may make a configuration inconsistent, and thus no longer a configuration at all). The macro above ensures that if we compare two configurations, $S_1$ and $S_2$, $t_{2,c}$ obtains the value $x_1$ in $S_1$ and the value $x_2$ in $S_2$, then we will prefer *over this criterion (independently of other criteria)* the configuration in which $t_{2,s}$ obtains the value which results in higher satisfaction. The macro thus conveys the idea that, whenever we are given two values of $t_{2,c}$, we prefer the one that we are more satisfied with.

Fuzzy relaxation of a quality constraint over a variable $v$ thus works by (i) removing the quality constraint on $v$, (ii) adding a fuzzy membership function $\mu(v)$ on $v$, (iii) interpreting $\mu(v)$ as the level of satisfaction with the value $v$, and (iv) adding a softgoal macro which generates preference relations that reflect the shape of $\mu$.

*3) Preferences from Softgoals:* Not all softgoals are macros used in fuzzy relaxation. A softgoal can be introduced in $\Delta$ even if we do now know the exact variable it refers to, or the exact range of values that it makes desirable. Very broad softgoals are allowed, such as

**s**($\tilde{p}_{17}$: Ambulances should quickly arrive at incidents)

where we write $\tilde{p}$ because the content of the softgoal is not a propositional variable as in a goal or task, since it is not clear from $\tilde{p}_{17}$ how to measure the time of arriving to incident scenes (e.g., does it begin at call reception, or at ambulance mobilization?) and it is not clear when a time to arrive at an incident counts as *quick*. There are two ways to *approximate* such softgoals: by refinement or by satisfaction functions, and both can result in preferences being added to $\Delta$.

When a softgoal is refined, it is treated just as any other requirement. E.g., we may define the following refinement of **s**($\tilde{p}_{17}$)

**q**($t_{7,e} \leq$ 15min);
**k**(**q**($t_{7,e} \leq$ 15min) $\to$ **s**($\tilde{p}_{17}$));
**k**($t_{7,e} = t_{1,c} + t_{2,c} + t_{3,c} + t_{4,e} + t_{5,e}$);

where we refined the softgoal with the quality constraint over the quantitative variable $t_{7,e}$ (i.e., if we satisfy **q**($t_{7,e} \leq$ 15min) then the if-then domain assumption above, with implication, tells us to deduce **s**($\tilde{p}_{17}$)), which is itself a function of times introduced before. We could find another refinement of **s**($\tilde{p}_{17}$), and have a preference between the two.

A softgoal can be approximated using a satisfaction function, and by proceeding in a similar way to fuzzy relaxation. Given **s**($\tilde{p}_{17}$), we decide which variable it refers to, and we assume it is $t_{7,e}$, such that **k**(**q**($t_{7,e} \leq$ 15min) $\to$ **s**($\tilde{p}_{17}$)). Secondly, we remove the softgoal **s**($\tilde{p}_{17}$). Thirdly, we define a function $\mu_1(t_{7,e})$, and we interpret the value given by $\mu_1(t_{7,e})$ as the level of satisfaction with the value $t_{7,e}$. Fourthly, we add a softgoal macro over $t_{7,e}$:

**s**( LET $x_1 \stackrel{\text{def}}{=}$ VAL($S_1, t_{7,e}$) AND $x_2 \stackrel{\text{def}}{=}$ VAL($S_2, t_{7,e}$);
IF $\mu_1(x_1) > \mu_1(x_2)$ THEN ADD
{**k**($t_{7,e} = x_1$), **k**($t_{7,e} = x_2$),
**k**($t_{7,e} = x_1$) $\succ$ **k**($t_{7,e} = x_2$)}
TO $\Delta$; )

This case is also illustrated with the softgoal **s**($\tilde{w}_1$) and function $\mu(t_x)$ in Figure 1(d).

*4) Use of Criteria for Ranking:* Given a set of criteria and roadmaps, two kinds of *decision rules* are needed: (i) to establish a ranking of configurations, from the most desirable to the least desirable, in a roadmap; (ii) to establish a ranking of roadmaps. RPAS, roadmap, and configuration concepts specify no particular decision rule. This is intentional, as no universal decision rule can be given: every decision rule gives one or more criteria priority over others, so that domain- and/or project-independent decision rules should be much more interesting. Note that the decision rule to apply will depend on how tradeoffs are resolved, through, among others, stakeholder negotiations, these concerns remain outside the scope of the general problem definition.

Various decision rules for configuration ranking can be used, such as one of the following:

R1: Rank highest the configuration $S$ if $S$ maximizes the quantitative variable $v$, and rank other configurations in a descending order by value of $v$ that each achieves: i.e., $S$ s.t. MAX($v$) iff $\forall S'$, $S'$ is a configuration and $max(\text{VAL}(S,v)) \geq max(\text{VAL}(S',v))$, where *max* returns the largest constant in a given set.

R2: Same as R1, but rank highest the configuration which minimizes $v$, i.e., rank from MIN($v$) to configuration which results in the highest value of $v$.

R3: Same as R1, augmented for the following condition: the chosen configuration should satisfy the maximal number of optional and preferred requirements, where in a preference relation $\phi \succ \psi$ or $\phi \succeq \psi$, $\phi$ is the preferred requirement.

R1 and R2 ignore preferences and optional requirements which are independent of $v$, and both give highest priority to the value of a chosen variable. In that case, the decision rule requires us to solve a single-objective optimization problem. The third rule is significantly different, as it has two high priorities, namely the value of the chosen quantitative variable and the number of optional and preferred requirements. The application of the third rule above requires the resolution of a multi-objective optimization problem.

The decision-rules that ranks roadmaps focuses on the properties of entire roadmaps. Maximization or minimization of a quantitative variable can now be extended over an entire roadmap, whereby we may ask that optimization satisfies constraints on efficiency of adaptations. E.g., the decision rule may be as follows:

R4: Choose roadmap $\mathcal{R}$ if (i) it maximizes the sum of values of the quantitative variable $v$ over all configurations in $\mathcal{R}$, under the constraints that (ii) $v$ must never fall below value $x$ in $R$, and (ii) no configuration in $\mathcal{R}$ differs from the configuration which immediately precedes it by more than $y$ requirements.

R4 is a decision rule pattern, where a concrete rule is obtained by setting $x$ and $y$ to constants. R4 illustrates that constraints need not be limited on individual configurations, but also on the structure of the roadmap, as it limits the number of changes that can be performed, which may reflect the necessity to conserve resources during adaptations.

It is important to note that the choice of a roadmap does not impose the *sequence* of configurations that the system should follow: adaptive behavior means that, e.g., if we adopt rule R4 above, the system will start from the configuration in which the domain assumptions are consistent with the conditions in the environment of the system. It will then adapt according to the satisfaction/failure of monitored requirements. It may be relevant to dynamically (at runtime) compute new configurations, the discussion of which we leave as an open issue.

## IV. Departures from Prior Work

The widely accepted general requirements problem definition is Zave & Jackson's [16] (ZJ hereafter), stating that RE should produce a specification ($S$) of a design of the system-to-be, which ensures that the system-to-be is consistent with domain assumptions ($K$) and together with them entails requirements ($R$): i.e., that $K, S \vdash R$. This definition emphasizes two properties: (i) *consistency*, in that the operationalization of requirements, i.e., $K \cup S$ must be consistent; and (ii) *achievement* of requirements, as $K \cup S$ are only acceptable if they are sufficient to deduce $R$.

For an example of ZJ problem and solution, cf., Figure 1(a). Assume that all goals there are the set $R'$, all domain assumptions $K'$, and all tasks $S'$. Given $R'$ and $D'$ initially, the ZJ problem says that we should find tasks $S \subseteq S'$ which satisfy a consistent subset $D \subseteq D'$ of domain assumptions, and a consistent subset of requirements $R \subseteq R'$. Any one column among $S_1$, $S_2$, and $S_3$ in Figure 1(c) is a consistent set of domain assumptions and tasks which satisfy a set of consistent goals (i.e., $\forall i,\ 1 \leq i \leq 3,\ S_i = K \cup S$), so that any one of these is a solution to the ZJ requirements problem. This can be verified by looking at the conflict and conditional relations in Figures 1(a)–(c).

There is in the ZJ problem formulation no information to define comparison criteria. We offered a revised definition of the requirements problem [8] (JMF hereafter), which has a richer ontology (to account for concepts that have established themselves in RE, such as, e.g., goals and softgoals), allows inconsistencies between alternative configurations (each of which satisfies ZJ conditions), introduces preferences to allow the comparison of configurations in terms of desirability, and we argued that logical consequence in early RE is more adequately described by a nonmonotonic and paraconsistent consequence relation $\mathrel{\vdash\mkern-9mu\sim}$ [7] than $\vdash$ from classical logic. These revisions make it impossible to synthesize the requirements problem as $K, S \vdash R$, since consistency and achievement become only a part of the conditions that a solution should satisfy. Note that alternative solutions are indistinguishable in ZJ because of the absence of comparison criteria.

Both the ZJ and JMF formulations of the requirements problem place considerable emphasis on properties which can be established from qualitative variables alone. With ZJ, the aim was to design one solution which is consistent and achieves requirements. With JMF, the aim was to design several configurations, all of which are consistent and achieve a threshold (i.e., all mandatory requirements), then proceed to decision-making, that is, compare configurations in terms of which preferred and optional requirements they satisfy, then choose one of them. Now, the aim is to choose roadmaps, whereby the configurations in a roadmap all are consistent and achieve requirements (as in ZJ), but also distinguish mandatory from optional requirements and can be compared in terms of preference (as in JMF). By satisfying Consistency and Qualitative threshold achievement, each configuration in a roadmap satisfies the ZJ conditions. By satisfying Conformity in addition, every configuration satisfies all properties that a candidate solution should satisfy in JMF. From there on, the departures of the configuration, adaptation requirement, roadmap, and RPAS concepts from ZJ and JMF are:

1) A configuration satisfies Quantitative threshold achievement, Dominance, and Minimality, in addition to Consistency, Qualitative threshold achievement, and Conformity.
2) Decision rules for the ranking of configurations and of roadmaps can be formulated as an optimization problem, where the aim is to optimize one or more quantitative variables. This was not of interest in ZJ, and was very limited in JMF.
3) There are new kinds of preferences and tradeoffs to consider. While a configuration may give an optimal value of a variable in one roadmap, we may prefer another roadmap, which fails to include that configuration, but ensures some suboptimal, but acceptable value of the variable over several configurations in the roadmap. Instead of focusing on the desirability of individual solutions, it is necessary to look at the desirability of roadmaps.

As the problem of designing, finding, and ranking roadmaps, RPAS reflects the consequences of adaptation behavior: a roadmap is a set of configurations, which need not have a predefined sequence, but adaptation requirements, while the system needs to be designed so as to be able to monitor requirements, and satisfy adaptation requirements by controlling its behavior, whereby it switches between the highest ranking, yet feasible configurations.

## V. Formalization

The aim in this section is to define the modeling language used throughout the preceding sections, then define additional tools necessary to present the formal definition of the configuration concept.

### A. Language

The set of requirements $\mathcal{L}$ is the union of four disjoint sets of formulas: simple requirements on propositional variables $\mathcal{L}^P$, simple requirements on quantitative variables $\mathcal{L}^N$, simple softgoals $\mathcal{L}^S$, and complex requirements $\mathcal{L}^C$.

Every **simple requirement on a propositional variable**, i.e., every $a \in \mathcal{L}^P$, satisfies the following BNF specification:

$$a ::= \mathbf{k}(p) \mid \mathbf{g}(p) \mid \mathbf{t}(p) \tag{V.1}$$

where $p$ is a symbol for a propositional variable. A requirement over a propositional variable can be a domain assumption ($\mathbf{k}$), a goal ($\mathbf{g}$), or a task ($\mathbf{t}$). The informal reading of the members of $\mathcal{L}^P$ is

- $\mathbf{k}(p)$: it is believed that $p$ is satisfied;
- $\mathbf{g}(p)$: it is desired that $p$ be satisfied;
- $\mathbf{t}(p)$: the execution of the task makes $p$ satisfied.

In the readings above, "can be" is replaced by "is" when the requirement is mandatory (see Complex requirements below).

Every **simple requirement on a quantitative variable**, i.e., every $b \in \mathcal{L}^N$, satisfies this BNF specification:

$$x ::= n \mid v \mid x+x \mid x-x \mid x \cdot x \mid x/x \mid x^x \tag{V.2}$$

$$\begin{aligned} y ::= &\, x > x \mid x < x \mid x = x \mid x \geq x \\ & \mid x \leq x \mid x \neq x \mid x \curvearrowright pdf \end{aligned} \tag{V.3}$$

$$b ::= \mathbf{k}(y) \mid \mathbf{q}(y) \mid \mathbf{t}(y) \tag{V.4}$$

where $n$ is a real number, $v$ is a quantitative variable, and *pdf* is a probability density function. The informal reading is:

- $\mathbf{k}(y)$: it is believed that the condition $y$ is satisfied;
- $\mathbf{q}(y)$: it is desired that the condition $y$ be satisfied;
- $\mathbf{t}(y)$: the execution of the task makes $y$ satisfied.

Again, "can be" is replaced by "is" above, when the requirement is mandatory.

**Simple softgoals** are kept outside $\mathcal{L}^P$ and $\mathcal{L}^N$ because the natural language statement in a softgoal (e.g., *High usability*, *High performance*), denoted $\tilde{p}$, refers to some desirable values of variables in such a way that it is not clear which exact values, or how the values of these variables can be observed or measured. Every $c \in \mathcal{L}^S$ satisfies:

$$c ::= \mathbf{s}(\tilde{p}). \tag{V.5}$$

An $\tilde{p}$ is called the content of a softgoal, and is a vague proposition. By being vague, $\tilde{p}$ is neither a propositional variable, nor a quantitative variable, nor a condition on quantitative variables. The symbol used is intentionally different from those inside domain assumptions, goals, quality constraints, and tasks.

A **complex requirement** relates simple requirements and/or softgoals, or is a mandatory or optional requirement. Every $h \in \mathcal{L}^C$ satisfies the BNF specification

$$d ::= a \mid b \mid c \tag{V.6}$$

$$e ::= \bigwedge_{i=1}^{m \geq 1} d_i \to d \;\Big|\; \bigwedge_{i=1}^{m \geq 2} d_i \to \bot \tag{V.7}$$

$$f ::= \mathbf{k}(e) \tag{V.8}$$

$$g ::= d \mid f \tag{V.9}$$

$$h ::= f \mid g^{\mathsf{M}} \mid g^{\mathsf{O}} \tag{V.10}$$

The specification of complex requirements introduces the convention that every complex requirement that has the implication connective is a domain assumption, as we assume that requirements or variables are related in refinement, conflict, or otherwise. Note that the quantitative refinement relation is specified as $\mathbf{k}(y)$, where $y$ is a condition in which a quantitative variable is equated to a function over quantitative variables.

We use lowercase letters of the Greek alphabet to denote generic requirements, members of $\mathcal{L} = \mathcal{L}^P \cup \mathcal{L}^N \cup \mathcal{L}^S \cup \mathcal{L}^C$. Uppercase Greek letters denote sets of requirements. Preferences over simple requirements and softgoals are kept in the separate set.

The set of **preferences** $\mathcal{P}$ includes all preference relations over simple requirements and/or softgoals; every $w \in \mathcal{P}$ satisfies the specification

$$w ::= d \succeq d \mid d \succ d \mid d \approx d. \tag{V.11}$$

Note from this specification that we do not allow preferences between complex requirements, and we not allow a preference to be mandatory or optional, to participate in a refinement, or in a conflict.

**Definition V.1.** *The language is the tuple $(P, \mathbb{R}, V, \tilde{P}, \mathcal{L}, \mathcal{P})$, where $\mathcal{L}$ the set of requirements which satisfy the BNF specification in Equations V.1–V.10 and $\mathcal{P}$ is the set of preferences which satisfy the BNF specification in Equation V.11, whereby the propositional variables in requirements and preferences come from the set $P$, numbers from the set $\mathbb{R}$ of real numbers, quantitative variables from $V$, and vague propositions from $\tilde{P}$.*

**Definition V.2.** *Any $\Delta \subseteq \mathcal{L} \cup \mathcal{P}$ is a requirements database.*

### B. Consequence Relation

We use the Techne consequence relation.

**Definition V.3.** *The **consequence relation** [7] $\vdash_{\tau}$ is such that, for $\Pi \subseteq \mathcal{L}$, $\phi \in \mathcal{L}$ and $x \in \{\phi, \bot\}$:*

- $\Pi \vdash_{\tau} \phi$ *if* $\phi \in \Pi$, *or*
- $\Pi \vdash_{\tau} x$ *if* $\forall 1 \leq n$, $\Pi \vdash_{\tau} \phi_i$ *and* $\mathbf{k}(\bigwedge_{i=1}^{n} \phi_i \to x)^{\mathbf{y}} \in \Pi$, *for* $\mathbf{y} \in \{\text{"empty"}, \mathsf{O}, \mathsf{M}\}$.

The consequence relation $\models_?$ is sound with regards to standard entailment $\vdash$ in classical propositional logic, but is incomplete in two ways: it only considers deducing positive atomic facts, and no ordinary proofs based on arguing by contradiction go through, thus being paraconsistent.

## C. Operationalization Functions

We have two operationalization functions, respectively for requirements over qualitative and over quantitative variables. The purpose of an operationalization function is, given a requirement $\phi$ and a set of requirements $X$, to return, if it exists, every set $Y \subseteq X$ which satisfies the following conditions. If $\phi$ is a requirement over a qualitative variable, then $Y$ includes only those domain assumptions and tasks which are necessary to deduce $\phi$. If $\phi$ is a requirement over a quantitative variable, then $Y$ includes only those domain assumptions and tasks which ensure that the condition in $\phi$ is satisfied, when all of the variables in that condition are replaced by the values that they are assigned by the requirements in $Y$.

To define operationalization functions, we define below the Select function and one useful set. We assume that $\mathcal{O} = \{\mathbf{k}, \mathbf{g}, \mathbf{q}, \mathbf{s}, \mathbf{t}\}$ and we denote $\wp(X)$ the powerset of $X$.

**Definition V.4.** *The select function*

$$\mathsf{Select} : \mathcal{O} \times \{\text{"empty"}, \mathbf{o}, \mathbf{M}\} \times \wp(\mathcal{L}) \longrightarrow \wp(\mathcal{L}) \quad (V.12)$$

*is defined as follows, for $\mathbf{x} \in \mathcal{O}$, $\mathbf{y} \in \{\text{"empty"}, \mathbf{o}, \mathbf{M}\}$, and $\Pi \subseteq \mathcal{L}$:*

$$\mathsf{Select}(\mathbf{x}, \mathbf{y}, \Pi) \stackrel{def}{=} \{\phi \mid \mathbf{x}(\phi)^{\mathbf{y}} \in \Pi\}. \quad (V.13)$$

To simplify notation, we use the following abbreviation:

$$\mathsf{Select}(\mathbf{x}, \mathbf{y}, \Pi) \equiv \Pi_{\mathbf{x}}^{\mathbf{y}}. \quad (V.14)$$

Given a set of expressions, the Select function returns its subset, in which all members have the labels $\mathbf{x} \in \mathcal{O}$ and $\mathbf{y} \in \{\text{"empty"}, \mathbf{o}, \mathbf{M}\}$.

**Definition V.5.** *Useful set: Let $\Pi \subseteq \mathcal{L}$ and $\mathbf{x} \in \mathcal{O}$:*

$$\mathsf{CON}(\Pi) \stackrel{def}{=} \{\Phi \subseteq \Pi \mid \Phi \not\models_? \bot \text{ and } \bigcup_{\forall \mathbf{x}} \Delta_{\mathbf{x}}^{\mathbf{M}} \subseteq \Phi\} \quad (V.15)$$

$\mathsf{CON}(\Pi)$ *is the set of all consistent subsets of $\Pi$, whereby every one of these subsets must include all mandatory requirements from $\Delta$.*

The reason why every $\phi \in \mathsf{CON}(\Pi)$ must include all mandatory requirements is that we require the set of mandatory requirments to be consistent and, as we will see below, to be included in all operationalizations, because an operationalization must not be inconsistent with mandatory requirements.

**Definition V.6.** *The **qualitative operationalization function***

$$\mathsf{Op} : \bigcup_{\forall \mathbf{x}, \mathbf{y}} \Delta_{\mathbf{x}}^{\mathbf{y}} \longrightarrow \wp\left(\wp\left(\bigcup_{\forall \mathbf{z}, \mathbf{w}} \Delta_{\mathbf{z}}^{\mathbf{w}}\right)\right) \quad (V.16)$$

*for $\mathbf{y}, \mathbf{w} \in \{\text{"empty"}, \mathbf{o}, \mathbf{M}\}$, $\mathbf{x} \in \{\mathbf{g}, \mathbf{q}, \mathbf{s}\}$, and $\mathbf{z} \in \{\mathbf{k}, \mathbf{t}\}$, is defined as follows:*

$$\mathsf{Op}\left(\phi \in \bigcup_{\forall \mathbf{x}, \mathbf{y}} \Delta_{\mathbf{x}}^{\mathbf{y}}\right) \stackrel{def}{=} \{\Pi \in \mathsf{CON}(\Delta) \mid \Pi \models_? \phi$$

$$\text{and } \Pi \setminus \{\phi\} \subseteq \bigcup_{\forall \mathbf{z}, \mathbf{w}} \Delta_{\mathbf{z}}^{\mathbf{w}},$$

$$\text{and } \nexists \Phi \subset \Pi, \ \Phi \models_? \phi\}. \quad (V.17)$$

*Every member of the set $\mathsf{Op}(\phi)$ is a minimal consistent set of tasks and domain assumptions that is sufficient to operationalize the goal, quality constraint, or softgoal $\phi$. Informally, $\mathsf{Op}(\phi)$ tells us all ways in (i.e., subsets of) $\Delta$ of satisfying a goal, quality constraint, or softgoal, which are consistent with all mandatory requirements.*

There can be cases when $\phi$ is a requirement over quantitative variables and is satisfied, but the function $\mathsf{Op}$ finds no operationalizations thereof. This happens if there are no qualitative refinements of $\phi$. We use the quantitative operationalization function to identify requirements which satisfy an $\phi \in \mathcal{L}^N$.

**Definition V.7.** *The **quantitative operationalization function***

$$\mathsf{Ov} : (\mathcal{L}^N \cap \Delta) \longrightarrow \wp\left(\wp\left(\bigcup_{\forall \mathbf{z}, \mathbf{w}} \Delta_{\mathbf{z}}^{\mathbf{w}}\right)\right) \quad (V.18)$$

*for $\mathbf{y}, \mathbf{w} \in \{\text{"empty"}, \mathbf{o}, \mathbf{M}\}$, $\mathbf{x} \in \{\mathbf{k}, \mathbf{q}, \mathbf{t}\}$, and $\mathbf{z} \in \{\mathbf{k}, \mathbf{t}\}$.*

Let $\forall i$, $n_i \in \mathbb{R}$, $v_i \in V$, where every $v_i$ is a variable which appears in $\phi$ and there are $m$ such variables in $\phi$, i.e., $1 \leq i \leq m$. $\mathsf{Ov}$ is defined as follows.
$\Pi \in \mathsf{Ov}(\phi \in \mathcal{L}^N)$ if and only if:
1) $\Pi \in \mathsf{CON}(\Delta)$, i.e., $\Pi$ is consistent,
2) $\{\mathbf{x}(v_i = n_i)^{\mathbf{y}} \in \mathcal{L}^N \mid 1 \leq i \leq m\} \subseteq \Pi$, i.e., $\Pi$ includes a set of requirements over the $m$ quantitative variables in $\phi$,
3) $\forall 1 \leq i \leq m$, one or more of the following hold
   a) $\mathbf{x}(v_i = n_i)^{\mathbf{y}} \in \bigcup_{\forall \mathbf{z}, \mathbf{w}} \Delta_{\mathbf{z}}^{\mathbf{w}}$, i.e., $\mathbf{x}(v_i = n_i)^{\mathbf{y}}$ is among domain assumptions and tasks in $\Delta$,
   b) $\exists \Phi \in \mathsf{Ov}(\mathbf{x}(v_i = n_i)^{\mathbf{y}})$ s.t. $\Phi \subseteq \Pi$, i.e., a subset of $\Pi$ is a quantitative operationalization of $\mathbf{x}(v_i = n_i)^{\mathbf{y}}$,
   c) $\exists \Phi \in \mathsf{Op}(\mathbf{x}(v_i = n_i)^{\mathbf{y}})$ s.t. $\Phi \subseteq \Pi$, i.e., a subset of $\Pi$ is a qualitative operationalization of $\mathbf{x}(v_i = n_i)^{\mathbf{y}}$,
4) *and if we replace every variable $v_i$ in $\phi$ with the constant $n_i$, then the condition in $\phi$ holds.*

*Every member $\Pi$ of $Ov(\phi)$ is a consistent set which includes all requirements which (i) are operationalized, (ii) assign values to every quantitative variable $v_1, \ldots, v_m$ in $\phi$, and (iii) assign such constants $n_1, \ldots, n_m$ to the variables $v_1, \ldots, v_m$ that the functional relation specified over these variables in $\phi$ holds. $Ov(\phi)$ tells us all combinations of requirements which assign such constants to all variables in $\phi$ that $\phi$ is satisfied.*

### D. Macros

Macros automatically change the requirements database when it already contains some specific requirements, or when specific values of quantitative variables are observed. There are softgoal macros and quantitative conflict macros.

Macros rely on the special *monitoring function* VAL. Given a set of requirements $X \subseteq \Delta$, and a quantitative variable $v$, if there are tasks and domain assumptions in $X$ which assign a value $n \in \mathbb{R}$ to $v$ (e.g., a task such that $\mathbf{t}(v = n)$ where $n$ is a constant), then $n \in \text{VAL}(X, v)$.

**Definition V.8.** *The **monitoring function***

$$\text{VAL} : \wp(\Delta) \times V \longrightarrow \wp(\mathbb{R}) \qquad (\text{V.19})$$

*is defined as follows, for $X \subseteq \Delta$, $v \in V$, $\mathbf{x} \in \{\mathbf{k}, \mathbf{q}, \mathbf{t}\}$, and $\mathbf{y} \in \{\text{"empty"}, \mathbf{o}, \mathbf{M}\}$:*

$$\begin{aligned}\text{VAL}(X, v) = &\{n \in \mathbb{R} \mid \exists Y \subseteq X \text{ s.t.} \\ &Y \in Ov(\mathbf{x}(v = n)^{\mathbf{y}}) \cup Op(\mathbf{x}(v = n)^{\mathbf{y}})\}\end{aligned}$$
(V.20)

$\text{VAL}(X, v)$ is a set of all constants that are assigned through qualitative or quantitative operationalizations in $X$ to $v$.

The **quantitative conflict macro** applies to any quantitative variable $v$ for which $\text{VAL}(\Delta, v) \neq \emptyset$. As long as $|\text{VAL}(\Delta, v)| \geq 2$, the macro adds a conflict relation between every pair of assignments of different values to $v$. E.g., if $\text{VAL}(\Delta, v) = \{3, 7\}$, this means that there is at least one operationalization which assigns the constant 3 to $v$, and at least one other operationalization which assigns the constant 7 to $v$. The macro will add two quality constraints, $\mathbf{q}(v = 3)$ and $\mathbf{q}(v = 7)$ to $\Delta$, and will add the mandatory conflict $\mathbf{k}(\mathbf{q}(v = 3) \wedge \mathbf{q}(v = 7) \to \bot)^{\mathbf{M}}$.

**Definition V.9.** *For every quantitative variable $v \in V$ such that $\text{VAL}(\Delta, v) \neq \emptyset$, there is the following **quantitative conflict macro**:*

```
k( IF |VAL(Δ, v)| ≥ 2 THEN ∀x₁, x₂ ∈ VAL(Δ, v) ADD
      {q(v = x₁), q(v = x₂),
         k(q(v = x₁) ∧ q(v = x₂) → ⊥)ᴹ}
   TO Δ; ).
```

The purpose of the quantitative conflict macro is to ensure that no quantitative variable obtains two different assignments in one configuration.

The **softgoal macro** applies to any quantitative variable $v$ for which $\text{VAL}(\Delta, v) \neq \emptyset$ and for which there is a satisfaction function $\mu(v)$. This is the case when there is in $\Delta$ an operationalization of a requirement which assigns a value to $v$ and a domain assumption which defines a variable $v'$ as the output of the satisfaction function on $v$, i.e., $\exists \mathbf{k}(v' = \mu(v)) \in \Delta$. The softgoal macro automatically adds preference relations between sets of requirements which operationalize requirements that assign constants to $v$, whereby the preferences are defined to reflect the values returned by $\mu$.

**Definition V.10.** *For every quantitative variable $v \in V$ such that $\text{VAL}(\Delta, v) \neq \emptyset$ and $\exists \mathbf{k}(v' = \mu(v)) \in \Delta$, where $\mu$ is informally interpreted as a satisfaction function, there is the following **softgoal macro**:*

```
s( ∀x₁, x₂ ∈ VAL(Δ, v)
   IF μ(x₁) = μ(x₂) THEN ADD
      {k(v = x₁), k(v = x₂),
         k(v = x₁) ≈ k(v = x₂)}
   TO Δ;
   IF μ(x₁) > μ(x₂) THEN ADD
      {k(v = x₁), k(v = x₂),
         k(k(v = x₁) ∧ k(v = x₂) → ⊥),
         k(v = x₁) ≻ k(v = x₂)}
   TO Δ; ).
```

Two remarks are in order. Firstly, note that the preferences are added to reflect the values not of $v$ but of the satisfaction level $\mu(v)$ on $v$. When the satisfaction levels are different (the second case in the softgoal macro), then a conflict relation is introduced as well, which will ensure that a configuration does not satisfy two assignments of different constants $x_1$ and $x_2$ to $v$ whereby these different assignments are not equally preferred (i.e., $\mu(x_1) \neq \mu(x_2)$). Secondly, note the difference between the definition of the softgoal macro above, and the macro we used earlier (cf., §III-D2): there, the macro was simpler because if made two assumptions which we do not make above, namely, (i) we assumed that we already knew the configurations, and (ii) we assumed that each configuration was such that the variable $v$ was assigned exactly one constant in each of the configurations.

### E. Configuration Concept

**Definition V.11.** *A **configuration** $S$, defined from the requirements database $\Delta$, is a set*

$$S \subseteq \bigcup_{\forall \mathbf{x}, \mathbf{y}} \Delta_{\mathbf{x}}^{\mathbf{y}}, \text{ for } \mathbf{x} \in \{\mathbf{k}, \mathbf{t}\}, \mathbf{y} \in \{\text{"empty"}, \mathbf{o}, \mathbf{M}\} \quad (\text{V.21})$$

*of domain assumptions and tasks, which satisfies the following properties:*

1) *Consistency: $S \not\vdash \bot$;*
2) *Qualitative threshold achievement:*
   $\forall \phi \in \Delta_{\mathbf{z}}^{\mathbf{M}}, \mathbf{z} \in \{\mathbf{g}, \mathbf{s}\}, \exists \Pi \in Op(\phi)$ *such that* $\Pi \subseteq S$;
3) *Quantitative threshold achievement:*
   $\forall \phi \in \Delta_{\mathbf{q}}^{\mathbf{M}}, \exists \Gamma \in Ov(\phi)$ *such that* $\Gamma \subseteq S$;
4) *Conformity: $\Delta_{\mathbf{k}}^{\mathbf{M}} \cup \Delta_{\mathbf{t}}^{\mathbf{M}} \subseteq S$;*

5) *Dominance:* $\nexists S'$ s.t. both $S$ and $S'$ satisfy conditions 1–4 above, $S \subset S'$, and $\exists \Delta_{\mathbf{x}}^{\mathbf{o}} = S' \setminus S$, s.t. $\Delta_{\mathbf{x}}^{\mathbf{o}} \neq \emptyset$ and $\mathbf{x} \in \{\mathbf{k}, \mathbf{t}\}$;
6) *Minimality:* $\nexists S'$ such that $S'$ satisfies the conditions 1–5 above and $S' \subset S$.

The Threshold achievement property requires that a configuration satisfies all mandatory goals, quality constraints, and softgoals. Satisfaction depends on the presence of operationalizations in $S$, for each of the mandatory goals, quality constraints, and softgoals.

The Conformity property asks that all strict domain assumptions are not violated and all mandatory tasks are executed. The Achievement and Conformity properties ensure that the configuration satisfies all that *must* be satisfied.

According to the Dominance property, every configuration will be maximal with regards to optional requirements. This property formalizes the idea of the optional relation, as holding on requirements which are desirable to satisfy, but can be violated. A configuration should include as many defeasible domain assumptions and as many optional tasks, up to the point at which adding any further defeasible domain assumptions and/or optional tasks violates the Consistency, Qualitative and Quantitative threshold achievement, Conformity, or Minimality properties. The Dominance condition ensures that every configuration is Pareto efficient with regards to optional requirements, as this condition makes it impossible to add optional domain assumptions and tasks to any $S$ and still ensure that $S$ is a configuration.

The Minimality property requires that a configuration includes only the domain assumptions and tasks which are needed to satisfy exactly the Consistency, Threshold achievement, Conformity, and Dominance properties.

*F. Related Work & Discussion*

We have illustrated throughout the paper how the contributions here depart from prior efforts in the understanding of the requirements problem and solution concepts. The roadmap concept is inspired by our discussion of the role of contexts, time, and resources in the requirements problem for adaptive systems [10]. We have used Techne [7] here as a starting point and extended its syntax to allow requirements which place constraints on quantitative variables. This has resulted in a more general treatment of operationalization, as we have both qualitative operationalization and quantitative operationalization functions. As in the case of Techne, the formalism here does not provide a visual syntax, and is thus not itself a model language for early requirements in the same sense that, e.g., KAOS goal models [4] and *i*\* actor models [15] are. In this sense, same remarks apply as those that we had made when comparing Techne to KAOS, Tropos, and *i*\*. Two limitations of Techne are overcome here. Firstly, we can make explicit the constraints on quantitative variables. Secondly, Techne was informally criticized that it is blind to the distinction between stable facts and defeasible information: this is not the case, as optional domain assumptions here behave like defeasible information, while mandatory domain assumptions act like stable facts.

In the rest of this section, we look at how the language presented in this paper can be used to model information that was recognized as crucial in the research into the relaxation of requirements [9], [14], [2], the evaluation of their partial satisfaction [9], and the monitoring and control of requirements [5], [12]. The aim is not to suggest the language here as a general modeling language, but to show that it covers the main ideas presented in relation to the said topics.

*Fuzzy Relaxation.* Baresi et al. associate every fuzzy operator with a predefined membership function. For example, if there is a quality constraint $\mathbf{q}(v < 6\text{hrs})$ here (a goal $G(v < 6\text{hrs})$ for them, as they allow quantitative variables in goals), then relaxing it would amount to replace it with $\mathbf{q}(v <_f 6\text{hrs})$ (in their notation, $\mathcal{G}(v <_f 6\text{hrs})$), where $\leq_f$ is a fuzzy operator. Their interpretation of $v <_f 6\text{hrs}$ is that there is a fuzzy membership function $\mu$ which returns the level of satisfaction as a function of $v$, and the shape of $\mu$ is predefined (for $<_f$, it is positive and constant until $v = 6$, then decreases up to the satisfaction value 0 for some $v > 6$). The operator $<_f$ can be defined in our language by reproducing, in a domain assumption a function which has the form of the fuzzy membership function for $<_f$, as defined by Baresi et al. We can define as follows a macro which takes a fuzzy goal of the form $\mathcal{G}(v <_f n)$, with $n \in \mathbb{R}$, and transforms it into requirements that can be added to our requirements database $\Delta$:

```
∀𝒢(v <_f n) WHERE v ∈ V AND n ∈ ℝ
ADD {k(v' = μ(v))} TO Δ, AND APPLY THE
SOFTGOAL MACRO ON Δ and v,
```

where $v'$ is a quantitative variable, i.e., $v \in V$, and $\mu$ is a function which is defined according to the function pattern defined by Baresi et al. for the fuzzy operator $<_f$. It is straightforward to define similar macros for all other fuzzy operators defined by Baresi et al.

Baresi et al. also define binary connectives, such as fuzzy conjunction. These can be defined here as well, as functions of those variables, which are defined using fuzzy membership functions. Each binary fuzzy operator gives one function specified in a domain assumption and using an approximation relation. In the formalism from Baresi et al., one way to define fuzzy conjunction $\wedge_f$ between two variables $v_1$ and $v_2$, each of which has an accompanying fuzzy membership function $\mu(v_1)$ and $\mu(v_2)$, is as follows: $\mu(v_1 \wedge_f v_2) = \mu(v_1) \cdot \mu(v_2)$. In our language, $\mu(v_1)$ and $\mu(v_2)$ give satisfaction levels. The fuzzy conjunction connective between two quality constraints, respectively over variables $v_1$ and $v_2$ is introduced in $\Delta$ as the domain assumption $\mathbf{k}(v_3 = \mu(v_1) \cdot \mu(v_2))$, where $v_3$ is the joint level of satisfaction over variables $v_1$ and $v_2$.

*Probabilistic Relaxation.* To handle idealistic requirements,

Letier & van Lamsweerde [9] suggest the association of probability estimates to constraints on quantitative variables. This is what we allow in the language, and this has been illustrated earlier (cf., §III-D1).

*Monitoring & Control Variables.* It is on the basis of the ontology for requirements that we identify controlled variables: if a variable appears in a task, it is a controlled variable. Any variable can be monitored, but all variables in goals, quality constraints, and domain assumptions need to be monitored.

*Adaptation Rules,* or equivalently, reconciliation tactics [5], adaptive goals [2], adaptivity mechanisms [13]. We have not discussed how they ought to be specified or actually implemented, but we have suggested how they can be identified. The roadmap adopted for a system gives a set of configurations. It consequently indicates what changes between two configurations, i.e., which requirements are dropped, which become optional, which others become mandatory, and so on. Differences between every two consecutive configurations in the roadmap specify the effect that adaptation rules should have, regardless of how they are specified or implemented.

*Limitations.* While it may be possible to specify time and temporal constraints by having a quantitative variable, the values of which are defined by a clock and to map every configuration to a temporal interval, this would clearly not be to a convenient way to model temporal constraints on roadmaps. More appropriate in this respect may be requirements modeling languages built on top of linear temporal logic [4] or branching temporal logic [14]. It is, however, not clear at all how requirements models built with such languages relate to roadmaps, that is, are these requirements models describing constraints on a single configuration, or on parts of roadmaps. Techne has already been criticized for ignoring the notion of agent, and the fact that requirements belong to individuals which may thus be modeled as agents. We continue to ignore them here, mainly because they can be introduced in a straightforward manner, yet would complicate notation and presentation, without adding much to the main purpose of this paper. To add agents, consider first why they need to be added. If the aim is to help figure out who needs to negotiate requirements conflicts, then assume there is a set of identifiers for agents, and add a function which returns, for every requirement, one or more agents who agree on it. If the aim is to assign responsibilities of tasks to agents, then assume a set of identifiers for these agents, and define a function which maps every task to one or more agents responsible for its execution. Either of the two uses of agents has no influence on the structure of the roadmap problem, other than suggesting that RE does involve negotiation and the assignment of responsibilities.

## VI. CONCLUSIONS, LIMITATIONS & OPEN ISSUES

We have presented two results. Firstly, by defining the configuration, adaptation requirement, and roadmap concepts, we have suggested a precise definition of the requirements problem for adaptive systems. We have related these concepts to the key notions from RE of adaptive systems, namely, monitoring, control, adaptation requirements, probabilistic and fuzzy relaxation. This led us to argue that there are fundamental differences between Zave & Jackson's and our conceptions of the requirements problem, and the requirements problem for adaptive systems, its solution concepts, and the decision rules used to rank solutions. Secondly, we have used a simple modeling framework throughout the paper, based on Techne, and which serves as a proto-framework, being an illustration of features needed in future requirements modeling languages for early RE of adaptive systems.

In addition to limitations, there is a number of interesting open issues. The shift to configurations and roadmaps suggests that research into extended planning and single- and multi-objective optimization may be a source for further advances in RE frameworks and tool support. It is necessary to look into how the roadmap requirements problem relates to mixed-integer programming, and mixed-variable programming, how to make adaptation rules on the basis of roadmaps, what decision rules may be relevant for decision-making in RE. Work is needed on the integration of probabilistic and fuzzy relaxation, monitoring and control within *practical* modeling languages, for early and late requirements phases. This paper will hopefully inform these future efforts.